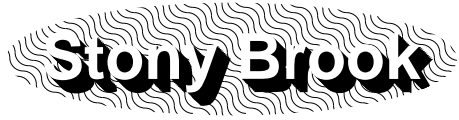

# THE N=2(4) STRING IS
# SELF-DUAL N=4 YANG-MILLS

W. Siegel[1]

*Institute for Theoretical Physics*
*State University of New York, Stony Brook, NY 11794-3840*

## ABSTRACT

N=2 string amplitudes, when required to have the Lorentz covariance of the equivalent N=4 string, describe a self-dual form of N=4 super Yang-Mills in 2+2 dimensions. Spin-independent couplings and the ghost nature of SO(2,2) spacetime make it a topological-like theory with vanishing loop corrections.

---

[1] Work supported by National Science Foundation grant PHY 90-08936.
Internet address: siegel@dirac.physics.sunysb.edu.

## 1. Background

Ooguri and Vafa [1] have shown that the N=2 open string [2] can be interpreted as self-dual Yang-Mills in 2+2 dimensions. (They did the same for closed strings and gravity. In this paper, we will consider only the Yang-Mills case for simplicity, but all statements about self-dual N=4 Yang-Mills can be applied analogously to self-dual N=8 supergravity.) Their treatment was based on the Yang formulation of self-dual Yang-Mills [3]. An equivalent formulation of self-dual Yang-Mills was given by Parkes [4]; the two forms are essentially different gauges which are related by a Lorentz transformation or a duality transformation [4]. Explicitly, defining the field strengths in terms of the covariant derivatives in SL(2)⊗SL(2) notation as

$$[\nabla_{\alpha\alpha'}, \nabla_{\beta\beta'}] = C_{\beta'\alpha'}F_{\alpha\beta} + C_{\beta\alpha}F_{\alpha'\beta'}$$

(where the $C$'s are the SL(2) antisymmetric symbols), the two formulations separate the self-duality conditions $F_{\alpha'\beta'} = 0$ as:

|  | constraints | field equation |
|---|---|---|
| Yang | $F_{+'+'}, F_{-'-'}$ | $F_{+'-'}$ |
| Parkes | $F_{+'+'}, F_{+'-'}$ | $F_{-'-'}$ |

This separation breaks the Lorentz invariance down to GL(2)=SL(2)⊗GL(1). The full Lorentz group transforms between the two formalisms, as does a duality transformation, which switches constraints with field equations. The Yang case is the analog of a temporal gauge, while the Parkes case is the analog of a light-cone gauge. The common constraint $F_{+'+'} = 0$ says that $\nabla_{\alpha+'}$ commute, so we can choose a gauge where $\nabla_{\alpha+'} = \partial_{\alpha+'}$. For the Yang case $F_{-'-'} = 0$ means $\nabla_{\alpha-'}$ is also pure gauge, while in the Parkes case $F_{+'-'} = 0$ says the curl of $A_{\alpha-'}$ with respect to $\partial_{\alpha+'}$ vanishes, so it is a gradient. We can then plug these solutions into the field equations; the result is:

|  | constraint solution | field equation |
|---|---|---|
| Yang | $\nabla_{\alpha+'} = \partial_{\alpha+'}$, $\nabla_{\alpha-'} = e^{-V}\partial_{\alpha-'}e^V$ | $\partial^\alpha{}_{+'}e^{-V}\partial_{\alpha-'}e^V = 0$ |
| Parkes | $\nabla_{\alpha+'} = \partial_{\alpha+'}$, $A_{\alpha-'} = \partial_{\alpha+'}V$ | $\Box V + i(\partial^\alpha{}_{+'}V)(\partial_{\alpha+'}V) = 0$ |

As typical for light-cone gauges, the Parkes solution is simpler than the nonpolynomial Yang solution (at least for perturbation theory). The actions are given by integrating these field equations with respect to the fields: The Yang case gives a Wess-Zumino action, while the Parkes case gives a cubic action.

The only nonvanishing tree amplitude in these self-dual Yang-Mills theories is the 3-point amplitude, which can be read from the lagrangian (or field equations): It is of the form $\mathcal{A}_{\alpha'\beta'} \sim k^\alpha{}_{(\alpha'} p_{\alpha\beta')}$ for momenta $k$ and $p$, where the Yang case gives $\mathcal{A}_{+'-'}$ and the Parkes case $\mathcal{A}_{+'+'}$. The only thing missing is the external line factors; however, these clearly do not correspond to those of 3 vectors, since the amplitude is a second-rank antisymmetric tensor. Thus, the theory appears to be Lorentz noncovariant. This is particularly clear in the Parkes formulation: Under the manifest GL(1) of the broken SL(2), the field transforms as $V_{-'-'}$, so the lagrangian $\frac{1}{2}V_{-'-'}\Box V_{-'-'} + ...$ is obviously not invariant under even this linear, unbroken subgroup of the Lorentz group. This does not imply that the Parkes formulation is less covariant than the Yang one, only that the noninvariance is more obvious. For example, the Parkes formulation has a clearer physical interpretation within perturbation theory: $V_{-'-'}$ represents the physical helicity $+1$ component of the original gauge vector. (The helicity $-1$ component has been killed, since it represented the anti-self-dual polarization.) However, there is an ambiguity in the spin interpretation of the N=2 string due to picture changing [5]. In the equivalent N=4 string [6], a Lorentz covariant analysis of the Ramond sector indicates the presence of spinors. (The N=4 algebra of the N=2 string had also been studied earlier in an attempt to analyze the hidden Lorentz structure [7].) In this paper we show how consideration of different spins is important in recovering Lorentz covariance; in particular, the world-sheet "equivalence" of different spins under picture changing ("spectral flow") emerges as the trivial structure of the theory under supersymmetry transformations. (The occurence of both bosons and fermions introduces further simplifications.)

## 2. Results

As suggested by the N=4 string, we consider supersymmetric self-dual Yang-Mills directly in SO(2,2) superspace. We solve the constraints for all N≤4, in formulations analogous to those of both Yang and Parkes. In the Parkes case all independent fields can be expressed conveniently in terms of a light-cone superfield. Lorentz invariance is still broken only to GL(2), since the light-cone superfield depends only on the $\theta_a{}^{+'}$ part of the anticommuting coordinates. ("$a$" is an index of SL(N), the internal symmetry.) The superfield equations contain no spinor derivatives, so the 3-point couplings have the same momentum dependence for any external particles, bosons or fermions. The superfield equations are also N-independent, taking the same form as for the purely bosonic case N=0. Another consequence of the lack of spinor derivatives in the field equations is that all loop amplitudes vanish [8]. (This should not be too surprising: Since all trees vanish for higher than 3-point, obtaining 1-loop graphs by sewing together two legs of tree graphs indicates all 1-loop graphs vanish except maybe vacuum bubbles and tadpoles, and supersymmetry kills vacuum bubbles and usually tadpoles.) Since the only nonvanishing contribution to perturbation theory is the local 3-point tree, these theories are in some sense topological.

The action now takes the form

$$\int d^4x \, d^N\theta^{+'} \, \tfrac{1}{2} V_{-'-'} \Box V_{-'-'} + \tfrac{1}{3} i V_{-'-'} (\partial^\alpha{}_{+'} V_{-'-'})(\partial_{\alpha+'} V_{-'-'})$$

so Lorentz invariance can be achieved only for N=4 (canceling the GL(1) charge of the lagrangian with that of the $d^N\theta^{+'} = (d_{+'})^N$ integration). The action has the same form as the light-cone superfield action for non-self-dual N=4 Yang-Mills [9], except that the antichiral terms are dropped. The covariant form of the action can be found by covariantization of the component expansion of the light-cone action:

$$L = G^{\alpha'\beta'} F_{\alpha'\beta'} + \psi_a{}^{\alpha'} \nabla_{\alpha\alpha'} \psi^{a\alpha} + \phi^{ab} \Box \phi_{ab} + \phi_{ab} \psi^{a\alpha} \psi^b{}_\alpha$$

Because of the SO(2,2) spacetime metric, the action has equal numbers of fields with positive and negative Hilbert space metric: The 6 scalars have SO(3,3) (=SL(4)) internal symmetry metric ($\phi^{ab}\phi_{ab} \equiv \tfrac{1}{2}\epsilon_{abcd}\phi^{ab}\phi^{cd}$), the spinors appear as $A\partial B$ with $A$ and $B$ real instead of complex conjugates, and the spin-1 fields appear as the anti-self-dual half of the $GF$ term of the usual first-order form $GF - \tfrac{1}{2}F^2$ of the Yang-Mills lagrangian. ($G$ is an independent field, $F$ is the field strength of the gauge vector.) Thus $G$ is a lagrange multiplier enforcing self-duality of Yang-Mills, and such linear lagrange multipliers are known to give ghosts [10]. However, the appearance of $G$ as propagating spin 1 is required by N=4 supersymmetry: Applying 4 supersymmetry transformations to the helicity $+1$ Yang-Mills gives the helicity $-1$ $G$. This also gives an interesting exception to the rule that coupling spin 1 to spin 1 requires that both spins be part of the same Yang-Mills: The field equation $\nabla^{\alpha\alpha'} G_{\alpha'\beta'} = 0$ implies $[F^{\alpha'\beta'}, G_{\alpha'\beta'}] = 0$, which is consistent for the anti-self-dual "field strength" $G$ because the field strength $F$ of $\nabla$ is self-dual ($F^{\alpha'\beta'} = 0$).

The appearance of ghosts was forced by the SO(2,2) metric. The SO(2,2) form of the usual spinor kinetic term has ghosts because the two chiralities of spinor (needed for the $\overline{\nabla}$) are independent representations of SL(2)⊗SL(2) instead of complex conjugates as in SO(3,1). The bosons then have ghosts because of supersymmetry. (This is true even for ordinary Yang-Mills, where $F^2 \approx A_{+-'} \Box A_{-+'}$ with $A_{\pm\mp'}$ both real.) This appearance of equal numbers of states of positive and negative Hilbert space metric is characteristic of the type of topological theories obtained by gauge fixing a zero lagrangian [11], where some of the supersymmetry acts as BRST. The vanishing of loops also maintains the superconformal symmetry SSL(4|4) at the quantum level. This enlarged spacetime symmetry has the interesting property that the internal and spacetime symmetries are the same, SL(4) (=SO(3,3)).

## 3. Calculations

The commutation relations for the covariant derivatives $\nabla_A = d_A + A_A$ of super Yang-Mills on shell in superspace are [12]:

$$\{\nabla_{a\alpha}, \nabla_{b\beta}\} = C_{\beta\alpha}\tilde{\phi}_{ab}, \quad \{\nabla^a{}_{\alpha'}, \nabla^b{}_{\beta'}\} = C_{\beta'\alpha'}\phi^{ab}$$

$$\{\nabla_{a\alpha}, \nabla^b{}_{\beta'}\} = \delta_a^b i \nabla_{\alpha\beta'}$$

$$[\nabla_{a\alpha}, \nabla_{\beta\beta'}] = C_{\beta\alpha}\tilde{\psi}_{a\beta'}, \quad [\nabla^a{}_{\alpha'}, \nabla_{\beta\beta'}] = C_{\beta'\alpha'}\psi^a{}_\beta$$

$$[\nabla_{\alpha\alpha'}, \nabla_{\beta\beta'}] = C_{\beta'\alpha'}F_{\alpha\beta} + C_{\beta\alpha}F_{\alpha'\beta'}$$

We then impose self-duality [13] by killing $F_{\alpha'\beta'}$ and other field strengths with only $\alpha'$ or lower $a$ indices:

$$F_{\alpha'\beta'} = \tilde{\psi}_{a\alpha'} = \tilde{\phi}_{ab} = 0$$

For the N=4 case there normally also would be a constraint relating the scalars: $\tilde{\phi}_{ab} = \frac{1}{2}\epsilon_{abcd}\phi^{ab}$. However, this would kill all the remaining field strengths, which are related to $\phi^{ab}$ by spinor derivatives through the Jacobi identities, so we drop it. Instead, we have the reality of $\phi$, which already makes it an irreducible **6** of the internal SO(3,3). The vanishing of $\tilde{\phi}$ is sufficient to satisfy the nonabelian Jacobi identity on $[\phi, \tilde{\phi}]$ which would normally require the above relation of $\tilde{\phi}$ to $\phi$ in 3+1 dimensions [14]. Thus, our self-dual N=4 super Yang-Mills has the same number of degrees of freedom as the usual non-self-dual one, although the theories are different in their interactions (chiral for the self-dual case) and field representation.

We next regroup these conditions to draw an analogy to N=0:

$$[\nabla_A, \nabla_B\} = T_{AB}{}^C \nabla_C \quad for \quad \nabla_A = (\nabla_{\alpha+'}, \nabla^a{}_{+'}, \nabla_{a\alpha})$$

$$[\nabla_A, \nabla_B\} = 0 \quad for \quad \nabla_A = (\nabla_{\alpha-'}, \nabla^a{}_{-'})$$

$$[\nabla_A, \nabla_{b\beta}\} = T_{Ab\beta}{}^C \nabla_C \quad for \quad \nabla_A = (\nabla_{\alpha-'}, \nabla^a{}_{-'})$$

$$[\nabla_{\alpha(+'}, \nabla_{\beta-')}] = \{\nabla^a{}_{(+'}, \nabla^b{}_{-')}\} = [\nabla^a{}_{(+'}, \nabla_{\beta-')}] = 0$$

The first set of constraints is the analog of (and includes) $F_{+'+'} = 0$, and the solution is that these derivatives take their free values in an appropriate gauge:

$$\nabla_{\alpha+'} = \partial_{\alpha+'}, \ \nabla^a{}_{+'} = d^a{}_{+'}, \ \nabla_{a\alpha} = d_{a\alpha}$$

The second set is the analog of $F_{-'-'} = 0$, and the fourth set is the analog of $F_{+'-'} = 0$; they are solved as before for the Yang/Parkes versions:

$$Yang: \ \nabla_{\alpha-'} = e^{-V}\partial_{\alpha-'}e^V, \ \nabla^a{}_{-'} = e^{-V}d^a{}_{-'}e^V; \ d^{(a}{}_{+'}e^{-V}d^{b)}{}_{-'}e^V = 0$$

$$Parkes: \ A_{\alpha-'} = \partial_{\alpha+'}V, \ A^a{}_{-'} = d^a{}_{+'}V; \ d^{a\alpha'}d^b{}_{\alpha'}V + i(d^{(a}{}_{+'}V)(d^{b)}{}_{+'}V) = 0$$

The third set is chirality conditions for the $\theta$ coordinates:

$$d_{a\alpha}V = 0$$

At this point we stick to the Parkes version for simplicity, and because of its resemblance to the light-cone gauge. We now notice that all the interesting quantities appear at $\theta_a{}^{-'} = 0$ in $V$:

$$V \qquad\qquad \partial_{\alpha+'}V = A_{\alpha-'} \qquad\qquad \partial_{\alpha+'}\partial_{\beta+'}V = iF_{\alpha\beta}$$
$$d^a{}_{+'}V = A^a{}_{-'} \qquad\qquad \partial_{\alpha+'}d^a{}_{+'}V = i\psi^a{}_\alpha$$
$$d^a{}_{+'}d^b{}_{+'}V = i\phi^{ab}$$
$$d^a{}_{+'}d^b{}_{+'}d^c{}_{+'}V = \epsilon^{abcd}\psi_{d+'}$$
$$d^a{}_{+'}d^b{}_{+'}d^c{}_{+'}d^d{}_{+'}V = \epsilon^{abcd}G_{+'+'}$$

In fact, part of the above equations of motion just determine the stuff at $\theta_a{}^{-'} \neq 0$ in terms of that at $\theta_a{}^{-'} = 0$. So, without loss of generality, we can look at just the part of $V$ evaluated at $\theta_a{}^{-'} = 0$. The equation of motion for this part can be derived from the general $V$ by just hitting the above equation of motion with $d_a{}^\alpha d_{b\alpha}$: The result is

$$\Box V + i(\partial^\alpha{}_{+'}V)(\partial_{\alpha+'}V) = 0$$

which is N-independent, and thus the same as for N=0. Similar remarks then apply to the action, except for the integration over $\theta_a{}^{+'}$.

The form of the covariant action can be guessed from the light-cone (Parkes) one. We simply note which fields appear at which order in $\theta_a{}^{+'}$, and that the integral requires 4 $\theta$'s. This is the same as just checking that the GL(1) number cancels. Using also Yang-Mills gauge invariance, the result

$$L = G^{\alpha'\beta'}F_{\alpha'\beta'} + \psi_a{}^{\alpha'}\nabla_{\alpha\alpha'}\psi^{a\alpha} + \phi^{ab}\Box\phi_{ab} + \phi_{ab}\psi^{a\alpha}\psi^b{}_\alpha$$

gives the above Parkes-style action in the corresponding gauge, where $A_{\alpha-'} = \partial_{\alpha+'} V$, $\psi^a{}_\alpha = -i\partial_{\alpha+'} A^a{}_{-'}$, $\phi^{ab}$, $\psi_{a+'}$, and $G_{+'+'}$ are the only surviving components. Alternatively, we can use dimensional analysis and the broken internal GL(1) symmetry: The internal GL(4) of the field equations is broken to SL(4) by the action only by the one $\epsilon_{abcd}$ introduced through the $\theta$ integration, when we write the fields as $A_{\alpha\alpha'}$, $\psi^a{}_\alpha$, $\phi^{ab}$, $\psi^{abc}{}_{\alpha'}$, and $G^{abcd}{}_{\alpha'\beta'}$.

## ACKNOWLEDGMENTS

I thank Nathan Berkovits, Jim Gates, and Martin Roček for discussions.